\documentclass[aps,prd,showpacs,twocolumn,amssymb,superscriptaddress,floatfix]{revtex4}

\usepackage{graphicx}
\usepackage{longtable}

\usepackage{amsmath, amsthm, amssymb}
\usepackage{color}

\newcommand{\nl}{\nonumber\\ }
\newcommand{\pd}{\partial}

\def\be{\begin{eqnarray}}
\def\ee{\end{eqnarray}}

\def\rmd{{\rm d}}

\begin{document}
\title{Relaxation time ansatz and shear and bulk viscosities of gluon matter}

\author{A. S. Khvorostukhin}
\affiliation{ Joint Institute for Nuclear Research,  141980 Dubna, Russia}
\affiliation{Institute of Applied Physics, Moldova Academy of Science,
MD-2028 Kishineu, Moldova}

\author{V. D. Toneev}
\affiliation{ Joint Institute for Nuclear Research,  141980 Dubna,
Russia}

\author{D. N. Voskresensky}
\affiliation{National Research Nuclear University "MEPhI",
Kashirskoe sh. 31, Moscow 115409, Russia}
\affiliation{GSI, Helmholtzzentrum f\"ur
Schwerionenforschung GmbH, Planckstrasse 1, 64291 Darmstadt, Germany}

\begin{abstract}
Shear  and bulk  viscosity-to-entropy density ratios are
calculated  for pure gluon matter in a nonequilibrium mean-field
quasiparticle approach within the relaxation time approximation.
We study how different approximations used in the literature
affect the results for the shear and bulk viscosities. Though the
results for the shear viscosity turned out to be quite robust, all
evaluations of the shear and bulk viscosities obtained in the
framework of the relaxation time approximation can be considered
only as rough estimations.
\end{abstract}

\maketitle

\section{Introduction}
High-energy heavy-ion collisions at  RHIC and LHC energies have
shown evidence for a new state of matter characterized by very low
shear viscosity to entropy density ratio, $\eta/s$, similar to a
nearly ideal fluid (see ~\cite{Romatschke,Aamodt}). Lattice
calculations indicate that the crossover region between hadron and
quark-gluon matter has been reached in these experiments. On the
other hand,  for the pure gluon ${\rm SU}(3)$ theory lattice
calculations demonstrate the occurrence of the first-order phase
transition (see~\cite{BEG95,Pa09}). Recently, there appeared gluon
lattice data on ratios of the shear \cite{SN07,Ma07} and bulk
\cite{Me08,SN07} viscosity to the entropy density.

Among  various existing phenomenological approaches, quasiparticle
 models are used to reproduce results obtained in the lattice QCD
 (see~\cite{PKPS96,PC05,IST_05,KTV09,KTV10}). In the case of
 gluodynamics,
quasiparticle models  rely on the assumption that for a
temperature $T$ below the critical one, $T<T_c$, the system
consists of a gas of massive glueballs and, for $T>T_c$, the
system consists of deconfined  gluons. Perturbative estimates of
the shear and bulk viscosities in gluodynamics were performed in
\cite{NS05,Me08,CDDW} and nonperturbative evaluations were made in
\cite{BKR09,KTVglue1,BKR10,DA} within the quasiparticle models
with temperature-dependent masses in the relaxation time
approximation with some additional  ansatze resulting in
essentially different expressions for the bulk viscosities.

Here we continue to exploit the mean-field-based quasiparticle
model with parameters fitted in \cite{KTVglue1} to fulfill  the
modern lattice data. Using two possible ansatze for the collision
term in the relaxation time approximation we derive expressions
for the shear and bulk viscosities to entropy density ratios for
the pure gluon ${\rm SU}(3)$ theory at $T>T_c$ and compare the
results with the lattice data. Our results demonstrate ambiguities
in calculation of the bulk viscosity within the relaxation time
approach.

\section{The quasiparticle model}

We start with the expression for  the gluon  energy-momentum
density tensor in a relativistic mean-field model,
 \begin{eqnarray}\label{Tmunua}
T^{\mu\nu} =\int d\Gamma\, \frac{p^\mu p^\nu}{E}f(t,\vec{r};
E,\vec{p}) +g^{\mu\nu}B[m^2(\varphi)]~,
\end{eqnarray}
 obeying the conservation law
\begin{eqnarray}\label{consT}
\partial_\mu T^{\mu\nu}=0.
\end{eqnarray}
Here
\begin{eqnarray}
d\Gamma &=&d_g  \frac{d^3 {p}}{(2\pi)^3} ~,\quad  p^\mu =
\left\{E[m^2(\varphi)],\vec{ p}\right\}~,
 \nonumber
\end{eqnarray}
 for gluons  with a degeneracy factor $d_g=2(N^2_c - 1)=16$, $N_c =3$; $\varphi$
 is a mean field; summation over the repeated indices is implied. In this model
the quasiparticle energy is given by
 \be
E =\sqrt{\vec{p}^{\,\,2} +m^{2}(\varphi )}~.
 \label{qe}
\ee
We assume that the quasiparticle distribution function $f$
obeys
the kinetic equation
\begin{eqnarray}\label{kineq}
   E^{-1} p^\mu\pd_\mu f -\nabla E \nabla_{\vec{p}} f &=& \mbox{St} f,
\end{eqnarray}
where $\mbox{St} f$ is the collision term satisfying the
condition
 \be\label{consStE}
\int d\Gamma E \ \mbox{St} f =0.
\ee
 Applying (\ref{consT}) for $\nu
=0$ and using (\ref{Tmunua}) and (\ref{consStE}) we derive the
consistency condition for the effective bag constant
\begin{eqnarray}
\label{dBdphi}
  \frac{d  B}{d \varphi}&=&-\frac{dm}{d\varphi}\rho_s ,
\end{eqnarray}
where
\begin{eqnarray}
\rho_s = \int d\Gamma\,
  \frac{m}{E}\, f\,
\end{eqnarray}
is  the scalar density. Note that in these expressions the
quasiparticle energy is a functional of the distribution function
$E=E[f]$. In thermal equilibrium relation (\ref{dBdphi}) coincides
with the thermodynamical consistency condition derived in
\cite{GY95} and used in the model \cite{KTVglue1}.

The local equilibrium distribution function for a gluon is
 \be
 f^{\rm l.eq.} (p^{\mu}_{\rm l.eq.} ,x^{\mu} )=
 \left[e^\frac{p^{\mu}_{\rm l.eq.} u_{\mu}(t,\vec{r})}{T(t,\vec{r})}- 1
 \right]^{-1}~.
 \label{leqdf}
 \ee
Here $p^\mu_{\rm l.eq.} = (E^{\rm l.eq.},\vec{p})$,
$$E^{\rm
l.eq.}= \sqrt{\vec{p}^{\,2}+m^2 \{\varphi^{\rm
l.eq.}[T(t,\vec{r})]\}},
$$
and the four-velocity of the frame is $u^\mu \simeq [1,
\vec{u}(t,\vec{r})]$ for $|\vec{u}|\ll 1$.  Following
\cite{KTVglue1}, we use the pole  mass
 \be \label{mg}
   (m^{\rm l.eq.})^2\equiv m^2 [\varphi^{\rm l.eq.}(T)]=
\frac{N_c}{6} \ g^2(T) \ T^2
 \ee
 as the  gluon quasiparticle mass in local equilibrium.
 The temperature-dependent strong interaction coupling in the next
 to leading order is given by
 \be \label{coupl}
 g^2 (T)=\frac{48\pi^2}{11N_c[\ln (\frac{\lambda(T-T_s)}{T_c})^2 +
 \frac{102}{121}\ln \ln (\frac{\lambda(T-T_s)}{T_c})^2]}~.
 \ee
 All previous papers  exploited $g^2$ up to the leading order using
only the first term in the denominator of (\ref{coupl}). In
Ref.~\cite{KTVglue1},  working within this  leading order we
fitted the parameters $\lambda$ and $T_s$ to fulfill the new
lattice data  \cite{Pa09} for the reduced pressure, energy,
enthalpy and trace anomaly. However,  we found that in the
perturbative limit [at $\ln \ln (\frac{\lambda(T-T_s)}{T_c})^2\gg
1$] the correction including the double-logarithmic term yields a
larger contribution to $dm^2/dT^2$ than the leading logarithmic
term. Since one of our aims in the given paper is to improve our
fit in the perturbative limit, we keep the double-logarithmic
correction and tune parameters to reproduce thermodynamic
characteristics of the system. Our fit \cite{KTVglue1} with $g^2$
calculated up to leading order was performed with the parameters
$T_s/T_c=0.5853$, $T_c =265$~MeV, and $\lambda=3.3$. The new fit
with $g^2$ calculated up to subleading order yields
$T_s/T_c=0.015$, $T_c =265$~MeV, and $\lambda=1.53$. Our previous
fit of the reduced pressure, energy, enthalpy and trace
anomaly~\cite{KTVglue1}  proves to be almost unchanged. Therefore,
we do not redraw those figures here.

\begin{figure}[t]
\includegraphics[width=8.truecm,clip]{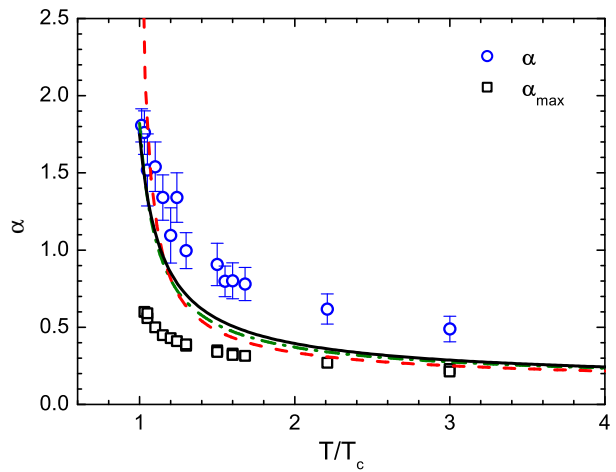}
\caption{(Color online) The effective running coupling $\alpha
(T)=g^2(T)/(4\pi)$ as a function of $T/T_c$. The points are the
lattice data \cite{Kaczmarek_07} (see explanation in the text).
The solid line is the result of our new fit~(\ref{coupl}). The
dash-dotted line is our (leading-order) fit  \cite{KTVglue1}. The
dashed curve is the (leading-order) fit of Ref.~\cite{BKR10} with
$T_s/T_c=0.73$, $T_c =271$~MeV, and $\lambda=4.3$.
 }
 \label{g2eff}
\end{figure}

Generally, the  running coupling constant $\alpha(T,r)=g^2 (T,r)/(4\pi)$
 depends on the temperature and distance and differs from the effective
one determined as $\alpha(T)=g^2 (T)/(4\pi)$
 following Eq.  (\ref{coupl}). The quadratic rise of
$\alpha(T=0,r)$ (see solid line in Fig. 6 of ~\cite{Kaczmarek_07})
is a nonperturbative effect that stems from the linear rising
string tension term in  the potential, while at small distances
the logarithmic weakening of the coupling is  visible and at
sufficiently small distances it reaches the perturbative behavior
(asymptotic freedom). At finite temperatures, $\alpha(T,r)$
follows this zero-temperature behavior to relatively large
distances, before Debye screening sets in, leading to a maximum
and a decrease at larger distances (see  Fig. 6 of
Ref.~\cite{Kaczmarek_07}). The value $\alpha$ is a
temperature-dependent coupling constant determined with the help
of the Debye screened potential at large distances
~\cite{Kaczmarek_07}. Thus, the behavior at large distances of the
coupling, which one may use to calculate thermodynamical
characteristics in quasiparticle models,  is not defined uniquely.
Therefore, in Fig.~\ref{g2eff} along with the  effective running
coupling constant $\alpha$ (circles)  the coupling constant
$\alpha_{max}$ (squares) is plotted to be defined by the maximum
in $\alpha(T,r)$~\cite{Kaczmarek_07}. As can be seen, both
estimates of the running constant noticeably differ near the
critical temperature and tend to coincide at high temperatures. In
Fig. \ref{g2eff}, we demonstrate also that the running constant
$\alpha (T)$ calculated up to the next to leading order with the
help of  Eq. (\ref{coupl})  in the given work (solid line) and
those calculated up to the leading order  with the parameters from
\cite{KTVglue1} (dash-dotted line) and from \cite{BKR10} (dashed
line)   deviate only little from each other, describing reasonably
the lattice data of Ref.~\cite{Kaczmarek_07}. Note that the
parametrization from Ref.~\cite{BKR10} is  fitted to the old
lattice data on thermodynamical quantities and does not describe
properly the new lattice data~\cite{Pa09}.

\section{Shear and bulk viscosities}

We define the shear, $\eta$, and bulk, $\zeta$, viscosities as
coefficients, entering into the variation of the energy-momentum
tensor in the local rest frame:
 \be
 \label{vis}
\Delta T_{ij}&=&-\zeta \
\delta_{ij}{\vec{\nabla}}\cdot\vec{u}+\eta \ W_{ij},\\ \nonumber
{\rm with} \quad W_{ij}&=&
\pd_iu_j+\pd_ju_i+\frac23\delta_{ij}\,{\vec{\nabla}}\cdot\vec{u}~.
 \ee
Here   Latin indices $1,2,3$ correspond to the spatial components.
Summation over the repeated  spatial indices is implied. To find
the shear viscosity, $\eta$,  we  set $i\neq j$ in (\ref{vis}). To
find the bulk viscosity, $\zeta$, we  substitute $i=j$ in
(\ref{vis}) and use the fact that $T_{ii}^{\rm l.eq}=3P^{\rm
l.eq}$.

The operation $\Delta$ in (\ref{vis}) needs explanation. According
to the work of Abrikosov and Khalatnikov \cite{AKh}, in
quasiparticle  Fermi liquid theory one usually  exploits the fact
that the following combination enters into the original Landau
collision term:
 \be \label{delta}
&&\delta \left(\sum_{a} E_a [f]\right)\cdot \left\{ f (E_1) f
(E_2)[1+ f (E_3)][1+ f (E_4)]\right.\nonumber
\\&&\left. -f (E_3) f (E_4) [1+ f(E_1)][1+ f (E_2)]\right\},
 \ee
where $E_a$ are the functionals of {\em the exact nonequilibrium
distribution function} $f_a$ and $a=1,2,3,4$. The term in curly
braces term is zero
 for $f^{\rm l.eq.} (E_a)$ given by
 \be
f^{\rm l.eq.}(E)=\left[e^\frac{p^\mu u_\mu (t,\vec{r})}{T
(t,\vec{r})}-1\right]^{-1} {\rm with}
 \, p^\mu =[E(\vec{p}),\vec{p}]\,,
 \label{leqdf1}
 \ee
where $E(\vec{p})$ is determined by Eqs. (\ref{qe}) and
(\ref{dBdphi}) and depends on the nonequilibrium distribution.
Thus, ${\rm{St}} f^{\rm l.eq.}
 (E )=0$. However, following \cite{KTVglue1}  the same collision
term should vanish also for local equilibrium, i.e., $\mbox{St}
f^{\rm l.eq.}(E^{\rm l.eq.})=0$. Let us demonstrate this by an
example of the $2\rightarrow 2$ processes, as they are described
by the four-momentum kinetic Kadanoff-Baym equation. Then the
local collision term renders \cite{IKV2}
\begin{eqnarray}\label{colFermi}
 &&C^{\rm loc.} = d^2 \int \frac{d^4{p_1}}{(2\pi)^4}
 \frac{d^4{p_2}}{(2\pi)^4}\frac{d^4{p_3}}{(2\pi)^4}|M|^2
\nonumber \\
&&\times  \delta^4
 \left(p + p_1 - p_2 - p_3\right)  A(x,p)A(x,p_1
)A(x,p_2 )A(x,p_3 ) \nonumber\\ &&\times\left\{ f
(x,p_{2})f(x,p_{3})
[1+ f(x,p_{1})][1+ f(x,p)]\right.\nonumber\\
&&\left. -[1+ f(x,p_{2})][1+ f(x,p_{3})]f(x,p_{1})f(x,p)\right\} ,
\end{eqnarray}
where $M$ is the  properly normalized  matrix element and $A=-2\mbox{Im}G^R$ is
the spectral function (density of states) that implicitly depends on the
corresponding distribution function $f$, $G^R$ is the retarded Green function,
 and $p_0$ and $\vec{p}$ are independent variables.
The Landau quasiparticle collision term is obtained from here
provided one sets $A[f]=2\pi \delta[p_0^2
-\vec{p}^{\,2}-m_0^2-\mbox{Re} \Sigma^R (p, [f])]$, where $m_0$ is
the free particle mass and $\Sigma^R$ is the retarded self-energy.
Here the local equilibrium distribution fulfills the   relation
\begin{eqnarray}
&& f^{\rm l.eq.}(x,p+q)[1+ f^{\rm
l.eq.}(x,p)]\nonumber\\
&&=[f^{\rm l.eq.}(x,p)- f^{\rm
l.eq.}(x,p+q)]f^{\rm l.eq.}(x,q),
\end{eqnarray}
where
 \be f^{\rm l.eq.}(x,p)=\left[\exp\left(\frac{p^\mu
u_\mu (t,\vec{r})}{T (t,\vec{r})}\right)-1\right]^{-1},
 \ee
i.e., in contrast to (\ref{leqdf}) and (\ref{leqdf1}), $p_0$ and
$\vec{p}$ are independent variables. With the help of this
relation we can see that the term in the curly braces in
(\ref{colFermi}) is zero independently of  values of $A$. Thus,
$C^{\rm loc.}=0$ for two distributions: $F\equiv iG^{-+}= A[f^{\rm
l.eq.}]f^{\rm l.eq.}$ and $F= A[f]f^{\rm l.eq.}$ (whereas the real
solution of the kinetic equation is $F=A[f]f$).  In our model $m$
does not depend on $p$, and the Landau quasiparticle collision
term is obtained provided one sets in (\ref{colFermi})  $A[f^{\rm
l.eq.}]=(\pi/E^{\rm l.eq.}) \delta(p_0 - E^{\rm l.eq.})$ (or
$A[f]=(\pi/E [f]) \delta(p_0 -E [f])$).  Note that in the Landau
quasiparticle kinetics one neglects memory effects; therefore,  we
indeed may use  the   local approximation in Eq. (\ref{colFermi}).

Then, returning to our quasiparticle model we can expand the distribution function
near $f^{\rm l.eq.}(E^{\rm l.eq.})$, i.e.  (\ref{leqdf}), performing variations as
 \be\label{deltaf}
 \delta f &=& f(E) -  f^{\rm l.eq.} (E^{\rm l.eq.}),
\\
\delta T^{\mu\nu} &=&T^{\mu\nu}[f(E)]-T^{\mu\nu}[f^{\rm
l.eq}(E^{\rm l.eq.})]\,,\nonumber
 \ee
i.e., with $\Delta =\delta$ in Eq. (\ref{vis}), and alternatively
we can introduce the operation  $\Delta$ as
 $\Delta=\widetilde\delta$, where
 \be\label{delFtilde} \widetilde\delta{f} &=&f
(E [f])-f^{\rm l.eq.} (E[f]),\\
  \widetilde\delta
T^{\mu\nu} &=&T^{\mu\nu}[f(E)]-T^{\mu\nu}[f^{\rm
l.eq}(E)]\,,\nonumber
 \ee
where $f^{\rm l.eq.} (E[f])$ is given by Eq. (\ref{leqdf1}).
Subtracting (\ref{deltaf})    from (\ref{delFtilde}) we obtain
the relation
 \be \label{connect}
\widetilde{\delta}f ={\delta}f-\left(\frac{\partial f}{\partial
E}\right)^{\rm l.eq.}\delta E \,.\ee Below we use both possible
choices of the $\Delta$ operation.

Performing variations in (\ref{Tmunua}) we find
 \be\label{delTik}
\Delta T_{ij}=\int d\Gamma\frac{p_i p_j}{E} \Delta f, \quad i\neq
j, \ee where we used the fact that after integration over angles
$\int d\Gamma p_ip_j (f/E^2)\Delta E =0$;  the diagonal terms
 \be\label{delTii} \Delta T_{ii}=\int
d\Gamma \frac{\vec{p}^{\,2}}{E}\Delta f -\int d\Gamma
\frac{\vec{p}^{\,2}}{E^2}f\Delta E  -3\Delta B,
\ee
where in the case $\Delta =\widetilde{\delta}$ the second and
the third terms vanish since $B=B(E)$  and $\widetilde{\delta}E=0$; and
\be\label{delT00}
\Delta T^{00}=\int d\Gamma E\Delta f.
\ee
To derive
(\ref{delT00}), we used the consistency relation (\ref{dBdphi}).

Since   in the local rest frame the viscosities are
pre-factors at small variations of the velocity,  in the linear
approximation used their values should not depend on how
variations were performed (using operation $\delta$ or
$\widetilde\delta$), provided all derivations are made without
additional assumptions. These values are usually computed with the
help of the local equilibrium distribution functions
(\ref{leqdf}).
 However, in practice one  uses additional approximations, e.g., great
simplification arises if one exploits the so-called relaxation
time approximation or more precisely  the relaxation time ansatz.
 Moreover, one can impose the so-called Landau-Lifshitz or somewhat different condition.
Below we show that the use of different ansatze can bring
significant differences in the resulting values of the kinetic
coefficients derived by means of the operations $\delta$ and
$\widetilde\delta$.

  In our previous work
\cite{KTVglue1}, to find shear and bulk viscosities of the gluon
and glueball matter we considered two relaxation time ansatze (see
Ref. \cite{CK10}) choosing the collision term as
 \be\label{St1} \mbox{St} f =-\,\delta f/\tau,
  \ee
and in a different form
\be\label{St2} \mbox{St} f =-\,\widetilde{\delta}
f/\widetilde{\tau}, \ee where $\tau$ and $\widetilde{\tau}$ are,
in general, different values and  in general they are energy- and
momentum-dependent quantities. All previous works
\cite{PC05,CDDW,BKR09,KTVglue1,BKR10,DA} used averaged values for
the gluon relaxation time.  In the latter case, as we show below,
the consideration can be performed consistently whereas for the
energy- and momentum-dependent relaxation time, problems can arise
[e.g., the two conditions (\ref{LLQ}) might be not fulfilled
simultaneously]. Thus, below we set ${\tau}$ and
$\widetilde{\tau}$ to be constant unless told otherwise.

 Additionally, for relativistic systems one usually uses the approach of
Landau and Lifshitz where $u^\mu$ is defined as the four-velocity
of the energy transport. Thus we require
\begin{align}
\Delta T^{i0}=0.
\end{align}
Then in the local rest frame the energy should
satisfy the Landau-Lifshitz condition \be\label{LL} \delta
T^{00}=0.
 \ee
 Also, performing all calculations at fixed exact
particle energy,
 i.e., with the help of the $\widetilde{\delta}$-operation, we can
 require fulfillment of the condition
 \be\label{LL1} \widetilde{\delta} T^{00}=0.
\ee

Thus, below we study three possibilities:

{\it ansatz I}: when the right-hand side (r.h.s.) of the kinetic
equation is presented in the form (\ref{St1}) and the
Landau-Lifshitz condition
 (\ref{LL}) is imposed,

 {\it ansatz II}: when the r.h.s. of the kinetic
equation is presented in the form (\ref{St2}) and the condition
(\ref{LL1}) is imposed, and

 {\it ansatz III}: when the r.h.s. of the kinetic
equation is presented in the form (\ref{St2}) but the
Landau-Lifshitz condition
 (\ref{LL}) is imposed.

\subsection{Relaxation time ansatz I}

  Let us first consider the kinetic equation with the r.h.s. in the form
(\ref{St1}) and use the Landau-Lifshitz condition
 (\ref{LL})  (ansatz I). Replacing $f\simeq f^{\rm l.eq.}
(E^{\rm l.eq.})$ [i.e., at $|\delta f|\ll  f^{\rm l.eq.} (E^{\rm
l.eq.})$],
 on the left-hand side (l.h.s.) of the kinetic equation (\ref{kineq}) we find
\be\label{lhs}
&&\left[ E^{-1} p^\mu\pd_\mu f -\nabla E \nabla_{\vec{p}} f\right]^{\rm l.eq.}
\\
&\simeq&-\left[\frac{f(1+f)}{2T E }\right]^{\rm l.eq.} p_kp_l
W_{kl} -\left[ Q \frac{f (1+f)}{T}\right]^{\rm
l.eq.}{\vec{\nabla}}\cdot\vec{u}\nonumber  .
 \ee
Introducing two terms
 \begin{eqnarray}
   \delta f&=& \delta f[W_{kl}]+\delta f
[{\vec{\nabla}}\cdot\vec{u}]
 \end{eqnarray}
and using (\ref{St1}) for the r.h.s. of (\ref{kineq}), we get
 \be\label{Feta} \delta
f[W_{kl}]=\left[\tau\frac{f(1+f)}{2T E }\right]^{\rm l.eq.} p_kp_l
W_{kl} , \ee and for $\delta f[{\vec{\nabla}}\cdot\vec{u}]$ we
find
 \be\label{delf}
\delta f [{\vec{\nabla}}\cdot\vec{u}]=\left[\tau Q \frac{f
(1+f)}{T}\right]^{\rm l.eq.}{\vec{\nabla}}\cdot\vec{u} .
 \ee
Here the equation-of-state-dependent $Q$ factor  (see
~\cite{KTV09,KTV10,KTVglue1}) is given by
\begin{eqnarray}
 \label{Qa} Q =
-\left\{\frac{\vec{p}^{\,2}}{3E}-c_s^2 \left[ E
-T\frac{\pd E}{\pd T} \right]
\right\}\,,
\end{eqnarray}
and
\be\label{cs}
c_s^2=\frac{\pd T_{ii}}{3\pd T^{00}}
 \ee
  is  the  speed of sound squared of the local equilibrium system.
To get (\ref{delf}), we used \be\label{Eder} \frac{\partial E^{\rm
l.eq.}}{\partial t} =\frac{d E^{\rm l.eq.}}{d \varphi^{\rm
l.eq.}}\frac{d \varphi^{\rm l.eq.}}{d T} \frac{\partial
T}{\partial t}\,. \ee Substituting (\ref{Feta}) into
(\ref{delTik}) and comparing with (\ref{vis}), we easily find the
expression  for the shear viscosity, \be \label{shear}
\eta=\frac{1 }{15T} \left\langle \tau \frac{\vec{p}^{\,4}}{E^2}\,
\right\rangle~,
 \ee
where for convenience we introduced the notation
\begin{eqnarray}
\left\langle \Phi(\vec{p}) \right\rangle =\int d\Gamma\,
\left[\Phi(\vec{p})f(1+f)\right]^{\rm l.eq.}.
\end{eqnarray}

Varying Eq. (\ref{dBdphi}) with respect to the mean field we obtain
\begin{eqnarray}\label{varphi}
  \delta \varphi&=&\Lambda\left(\frac{dm}{d\varphi}\right)^{\rm l.eq}\int d\Gamma
  \frac{m^{\rm l.eq.}}{E^{\rm l.eq.}}\, \delta f,
\end{eqnarray}
 with
\begin{eqnarray}
\label{defLambda}
  \Lambda^{-1}&=&\left(\frac{dm}{d\varphi}\right)^2_{\rm
  l.eq}\left(\int d\Gamma
  \left\{\frac{{m}^2}{{E}^3}f\right\}-\frac{\rho_s}{m}\right)^{\rm l.eq.}\nl
  &-&\left(\frac{d^2m}{d\varphi^2}\right)^{\rm l.eq}\rho_s^{\rm l.eq.} -\left(\frac{d^2  B}{d
  \varphi^2}\right)^{\rm l.eq.}.
  \end{eqnarray}
Calculating ${d^2  B}/{d \varphi^2}$ with the help of
(\ref{dBdphi}) we get
\begin{eqnarray}
\label{Lambda1}
  \Lambda^{-1}=\frac1T\left(\frac{dT}{d\varphi}\right)^2\left[ \frac{m}{T}\,
  \frac{dm}{dT}\,\langle1\rangle
-\left(\frac{dm}{dT}\right)^2\,\left<\frac{m^2}{E^2}\right>\right].
  \end{eqnarray}
Making use of $\delta m=(dm/d\varphi)\delta\varphi$ from Eqs.
(\ref{varphi}) and (\ref{Lambda1}) we have
 \be\label{dm}
 \delta m =\frac{T^2 \frac{dm}{dT}\int d\Gamma \frac{\delta
 f}{E}}{\langle1\rangle
 -Tm\frac{dm}{dT}\langle\frac{1}{E^2}\rangle}\,.
 \ee
Then from Eqs. (\ref{varphi}) and (\ref{delTii})  we find
\begin{eqnarray}
  \delta T_{ii}
  =\int \,d\Gamma\left\{\frac{\vec{p}^{\,2}-K}{E^{\rm l.eq.}}\right\}\delta f,
\end{eqnarray}
where
\begin{eqnarray}
 K&=&-Tm^2\,\frac{dm}{dT}
 \left<\frac{\vec{p}^{\,2}}{E^2}\right>\nl
  &\times&
 \left[ m\,\langle1\rangle
 -T\frac{dm}{dT}\,\left<\frac{m^2}{E^2}\right>\right]^{-1}\,.
\end{eqnarray}

Using (\ref{vis}) we obtain the final expression for the bulk
viscosity:
\begin{eqnarray}
\label{genzeta}
  \zeta&=&-\frac1T\left\langle
 \frac{\tau Q}{3E}\,\left(\vec{p}^{\,2} -K
 \right)\right\rangle .
  \end{eqnarray}
 Note that  for $\tau =\,$const the variable shift
\be\label{variable} Q\tau \to Q\tau +bE,
 \ee
 suggested in \cite{CK10} and then used  in \cite{BKR10,DA}
 to satisfy the Landau-Lifshitz condition (\ref{LL}),
is not required here, since Eq. (\ref{LL}) and the condition
 $\int d\Gamma E\,\mbox{St} f=0$, i.e.,
 \be\label{LLQ}
 \left\langle \tau Q E\right\rangle =0\,\quad {\rm and}\quad\left\langle
Q E\right\rangle =0\,,
  \ee
are fulfilled with $Q$ given by Eq. (\ref{Qa}).
This statement is easily checked with the help of the relation
\begin{eqnarray}
\label{cs2expr} (c_s^{\rm l.eq.})^2
=\frac13\frac{\langle
\vec{p}^{\,2}\rangle}{\langle E^2\rangle-Tm \frac{d m}{d
T}\,\langle1\rangle}\,.
 \end{eqnarray}
 To  reproduce the second relation (\ref{LLQ}) we used the result
 (\ref{delf}) and Eq. (\ref{St1}).

 Moreover, in the case of a momentum-dependent relaxation time and
 also  in the case of many particle
species with different values of the relaxation time, $\tau_a$,
 the  replacement $Q_a\tau_a \to Q_a\tau_a +bE_a$ does not generate new
solutions $\mbox{St}f_a =0$, provided   the collision term  is
presented within the relaxation time approximation. Thus, we note
that the replacement $Q_a\tau_a \to Q_a\tau_a +bE_a$  generates
new solutions $\mbox{St}f_a =0$ and thereby the  shift is
meaningful only if the exact  form of the local collision term
$\mbox{St}f_a$ is used.

 Nevertheless, compared to the results of other works,  performing  the
shift (\ref{variable}) one can  present Eq. (\ref{genzeta}) in the form
\be\label{genzetashift}
\zeta =-\frac1T\left\langle
 \frac{\tau Q}{3E}\,\left(\vec{p}^{\,2} -K-3\gamma E^2
 \right)\right\rangle ,
\ee
 with
  \be\label{gam} \gamma =\frac{\left\langle
\vec{p}^{\,2}\right\rangle -K\left\langle
1\right\rangle}{3\left\langle E^2\right\rangle}\,,
 \ee
 where
following (\ref{LLQ}) the last term in Eq. (\ref{genzetashift}) is
zero provided $\tau$ is constant. We note that, although  the
variable shift (\ref{variable}) has been performed, Eq.
(\ref{genzetashift}) does not yield the quadratic form exploited
in Refs. \cite{CK10,BKR10,DA}. Only for $m =\,$const do we recover
the quadratic form for the bulk viscosity \cite{G85}. We also note
that for a temperature-independent $g^2$, expression
(\ref{genzetashift}) would yield $\zeta =0$. This circumstance
gives additional justification to our inclusion of the
double-logarithmic correction to $g^2 (T)$ in Eq. (\ref{coupl}) in
order to correctly take into account the $m(T)$ dependence.

Now let us prove that expression (\ref{genzetashift}) is
nonnegative. For that we rewrite Eq. (\ref{genzetashift}) as
\begin{eqnarray}
\zeta&=&\frac{\tau}{9T}\left(m^2 -3c_s^2 T^2\frac{d m^2}{d
T^2}\right)\left(m^2 +K\right)\nl
&&\times\left[\left\langle\frac{1}{{E}^2}\right\rangle
-\frac{\langle{1}\rangle^2}{\langle  {E}^2\rangle}\right].
\end{eqnarray}
 From here for $m=\,$const we obtain
\begin{eqnarray}\label{mconst}
\zeta (m={\rm const})=\frac{\tau
m^4}{9T}\left[\left\langle\frac{1}{{E}^2}\right\rangle
-\frac{\langle{1}\rangle^2}{\langle  {E}^2\rangle}\right].
\end{eqnarray}
From Eq. (\ref{cs2expr}) and the condition $3c_s^2\leq 1$, it follows
 that $m^2\geq T^2\frac{dm^2}{dT^2}$ and $K\geq-m^2$.
Then making use of $\zeta (m={\rm const})\geq 0$ we find that
$\zeta\geq 0$.

Note that from relations (\ref{connect}), (\ref{delf}) and
(\ref{dm}) it  follows that the collision term (\ref{St1}) can be
expressed as
 \be\label{St1tauE}
 \mbox{St}f[\vec{\nabla} \cdot\vec{u}]=-\delta f[\vec{\nabla}\cdot
 \vec{u}]/\tau =\widetilde{\delta}f[\vec{\nabla}\cdot \vec{u}]
 /\widetilde{\tau}(E)
 \ee
with the energy-dependent relaxation time
 \be
\widetilde{\tau}(E)=\tau \left[1+\frac{Tm\frac{dm}{dT}\langle\tau
\frac{Q}{E}\rangle/(\tau EQ)}{\langle 1\rangle
-Tm\frac{dm}{dT}\langle\frac{1}{E^2}\rangle}\right]\,.
  \ee
 Expression (\ref{St1tauE}) (see the second equality)  has the same form as
(\ref{St2}), being postulated in ansatz II.  However, note that to
satisfy the Landau-Lifshitz condition in the form (\ref{LL}), we
assumed that $\tau$ is an averaged value and the relaxation time
$\widetilde{\tau}(E)$ determined by the second equality
(\ref{St1tauE}) proved to be a certain (momentum-dependent)
function depending on the constant value $\tau$. Thus,  starting
with (\ref{St1}) for $\tau=\,$const, we arrived at (\ref{St2})
with $\widetilde\tau=\widetilde\tau(E)$. Also in passing we
demonstrated that the collision term (\ref{St2}) [as well as
(\ref{St1})] vanishes in the local equilibrium state  [i.e., for
$f= f^{\rm l.eq.}(E^{\rm l.eq.})$].

\subsection{Relaxation time ansatz II}

Let us now assume  that the collision term is given by Eq.
(\ref{St2}) and perform all variations at a fixed nonequilibrium
energy, i.e., we apply the $\widetilde{\delta}$ operation to all
quantities. Then $\widetilde{\delta}E=0$ and
$\widetilde{\delta}B(E)=0$ and the distribution function $f^{\rm
l.eq.}(E)$ is given by (\ref{leqdf1}). Since we need to keep only
terms linear in $\partial_i u_k$ on the l.h.s. of the kinetic
equation, we may use there (\ref{leqdf}) instead of
(\ref{leqdf1}). Then using Eq. (\ref{lhs}) for the l.h.s. of
(\ref{kineq}) and Eq. (\ref{St2}) for the r.h.s. we find
 \be\label{part1}
\widetilde{\delta} f[W_{kl}]\simeq
\left[\widetilde{\tau}\frac{f(1+f)}{2T E }\right]^{\rm l.eq.}
p_kp_l W_{kl}
 \ee
 and
  \be\label{delf1}
\widetilde{\delta} f [{\vec{\nabla}}\cdot\vec{u}]\simeq
\left[\widetilde{\tau} Q \frac{f (1+f)}{T}\right]^{\rm
l.eq.}{\vec{\nabla}}\cdot\vec{u} .
 \ee
Applying here relation (\ref{connect}) and averaging the result
over angles we arrive at
 \be
 \widetilde{\delta}T^{ik}=\delta T^{ik},\,\,\,\mbox{for} \,\,\,
 i\neq k\,,
 \ee
  and also \be
 \widetilde{\delta}T^{i0}=\delta T^{i0}\,.
 \ee
By using the consistency condition,  relation (\ref{connect})
after partial integration gives
 \be \label{deltaTiiKap}\widetilde{\delta}
T_{ii} &=&\int d\Gamma
 \vec{p}^{\,2}\widetilde{\delta}f/E\nonumber \\
 &=&\int d\Gamma
 \vec{p}^{\,2}{\delta}f/E^{\rm l.eq.} ={\delta} T_{ii}\,.
 \ee
 Exploiting ansatz II  we replace the Landau-Lifshitz
condition (\ref{LL}) by condition (\ref{LL1}).
The last condition differs from (\ref{LL}) by the terms linear in
$\delta m\propto \vec{\nabla} \vec{u}$. We observe that  the
condition (\ref{LL1}) with $\widetilde{\delta}f$ satisfying
(\ref{delf1}) is fulfilled provided $\widetilde{\tau}$
is assumed to be a momentum-independent (averaged)
quantity since, as in ansatz I, Eq. (\ref{LL1}) is  reduced to
the same second condition (\ref{LLQ}) proved above for $Q$
from Eq. (\ref{Qa}).

Then we immediately recover  expressions for the shear viscosity,
\be\label{shear1} \eta=\frac{1 }{15T} \left\langle
\widetilde{\tau} \frac{\vec{p}^{\,4}}{E^2}\,  \right\rangle , \ee
and the bulk viscosity,
\begin{eqnarray}\label{genzeta1}
  \zeta &=&-\frac{1}{T} \left\langle
 \frac{\widetilde{\tau} Q\,\vec{p}^{\,2}}{3E}\right\rangle .
  \end{eqnarray}
For $m=\,$const, Eq. (\ref{genzeta1}) transforms to (\ref{mconst})
and reproduces the result of Ref. \cite{G85}.

  If for completeness of consideration,  the shift
\be\label{variable1} Q\widetilde{\tau}\rightarrow
Q\widetilde{\tau}+ \widetilde{b} E \ee
  is performed,  Eq. (\ref{genzeta1}) can be presented in the form
\begin{eqnarray}\label{genzeta1shift}
  \zeta =-\frac{1}{T} \left\langle
 \frac{\widetilde{\tau} Q\,\vec{p}^{\,2}}{3E}\right\rangle -
 \frac{1}{T} \frac{\left\langle\widetilde{\tau}
 EQ\right\rangle}{\left\langle\widetilde{\tau}
 E^2\right\rangle} \left\langle\frac{\widetilde{\tau}\vec{p}^{\,2} }{3}
 \right\rangle \,.
  \end{eqnarray}
The last term in Eq. (\ref{genzeta1shift}) vanishes if one uses
the averaged value of $\widetilde{\tau}$. Again, although the
variable shift  (\ref{variable1}) has  been done, Eq.
(\ref{genzeta1shift}) does not yield the quadratic form exploited
in Refs. \cite{CK10,BKR10,DA}. Also, one should stress once more
that the variable shift is meaningful only if the exact collision
term
 without memory terms is used.

Note that the result  (\ref{genzeta1}) follows from
(\ref{genzeta}) for $K=0$. Thereby, the final result
(\ref{genzeta1}) is nonnegative.

We assumed that  values $\tau$ and $\widetilde{\tau}$ entering
into the collision terms (\ref{St1})  and (\ref{St2}) are averaged
quantities (constants)  evaluated by means of the cross section.
In this case, we cannot distinguish between $\tau$ and
$\widetilde{\tau}$  and should set $\tau
=\widetilde{\tau}=\bar{\tau}$ with $\bar{\tau}$ being an averaged
relaxation time.  Thus, with such a simplified approach we are
unable to distinguish which expression, (\ref{genzeta}) or
(\ref{genzeta1}), is more preferable. Then for $\eta$  both
approaches (with $\Delta =\delta$
 and $\delta =\widetilde{\delta}$) result in the same expression.

 Exploiting  ansatz I in  \cite{KTVglue1} we further used an additional
ansatz, namely, in performing variations we did not vary
quantities which depend on the distribution function only
implicitly, such as $E$ and $B(E)$. Doing so we arrived at  Eq.
(\ref{genzeta1}). Thus, our results obtained in \cite{KTVglue1}
within ansatz I  are actually equivalent to those derived with
ansatz II in the given work. As follows from the comparison of
Eqs.  (\ref{genzeta}) and (\ref{genzeta1})  the approximation used
in \cite{KTVglue1} within ansatz I is  appropriate only for
$|K|\ll m^2$. The latter inequality holds for nonrelativistic
quasiparticles but it becomes a poor approximation in the
relativistic limit. As follows from Eq. (\ref{mg}) and Fig.
\ref{g2eff}, in our case the nonrelativistic approximation is not
applicable. Therefore, calculations with  one and the same value
$\bar{\tau}$ performed following Eqs.  (\ref{genzeta}) and
(\ref{genzeta1})  may differ significantly.

  Combining (\ref{connect}) and (\ref{Lambda1}), we find
\begin{eqnarray}\label{deltphiL}
 \delta\varphi &=&L \left(m\frac{dm}{d\varphi}\right)^{\rm l.eq.}
 \int d\Gamma\frac{1}{E^{\rm l.eq.}}\,\widetilde\delta
 f\,,\\
 L^{-1}&=&
 \Lambda^{-1}+\left\langle\frac1T\left(\frac{dm}{d\varphi}\right)^2
 \frac{m^2}{E^2}\right\rangle\nl &=&\left(\frac{dT}{d\varphi} \frac{m}{T^2}\,
 \frac{dm}{d\varphi}\right)^{\rm l.eq.}\,\langle1\rangle\,,
\end{eqnarray}
and
 \be
 \delta m=\frac{T^2}{\langle1\rangle} \frac{dm}{dT}\int d\Gamma
 \frac{\widetilde{\delta}f}{E}\,.
  \ee
Using now relations (\ref{connect}), and (\ref{delf1}) we are able
to express the collision term (\ref{St2}) in the form
 \be\label{St1E}  {\rm St}\, f[\vec{\nabla} \cdot\vec{u}] &=& -
 \frac{{\delta f}[\vec{\nabla} \cdot\vec{u}]}{{\tau}(E)},
\end{eqnarray}
with
\begin{eqnarray}
  {\tau}(E)=
 \widetilde{\tau}\left(1-\frac{Tm\frac{dm}{dT}\langle
 \frac{\widetilde{\tau}Q}{E}\rangle}
 {\widetilde{\tau}EQ\langle1\rangle}\right)\,.
\end{eqnarray}
Thus, if one assumes that Eq. (\ref{St2}) holds for
$\widetilde{\tau}\simeq\,$ const, then one recovers Eq.
(\ref{St1E}) with the momentum-dependent value of ${\tau}(E)$. In
contrast, if one assumes that Eq. (\ref{St1}) holds for
${\tau}\simeq\,$const, then one recovers Eq. (\ref{St1tauE}) with
the momentum-dependent value of $\widetilde{\tau}(E)$.

\subsection{Ansatz III}

It was  accepted in Ref. \cite{CK10} that ${\delta} T_{ii} =\int d\Gamma
 \vec{p}^{\,2}\widetilde{\delta}f/E^{\rm l.eq.}$
but the Landau-Lifshitz condition was   used in the form
(\ref{LL1}). Moreover, for simplification  the authors~\cite{CK10}
assumed  a specific form of the nonequilibrium distribution
function $f=e^{-E[f]/T[f]}$ (in the Boltzmann limit $f\ll 1$)
introducing variations of the temperature $\widetilde{\delta} T$.
The mentioned procedure results in a quadratic form for the bulk
viscosity.

Let us demonstrate how one can derive the same quadratic form for
$\zeta$ but slightly more consistently. If we ignored the
Landau-Lifshitz condition, then using the first line of
(\ref{deltaTiiKap}) and the kinetic equation within ansatz II
(\ref{St2}), we would recover expression (\ref{genzeta1}). On the
other hand,   using the second line of (\ref{deltaTiiKap}), we can
 expand the  energy-momentum tensor density near the local equilibrium
state and, therefore, exploit  Eqs. (\ref{vis}) and
(\ref{delTik})--(\ref{delT00})  for $\Delta =\delta$, taking,
nevertheless, the collision term in the kinetic equation  in the
form  (\ref{St2}).

We express (\ref{LL}) through $\widetilde\delta f$ using
(\ref{deltphiL}).
Then
\begin{eqnarray}
\label{LLKapusta}
 \delta T^{00}&=&\int d\Gamma E^{\rm l.eq.}\, \widetilde\delta
 f +\int d\Gamma \left(E\,\frac{\pd f}{\pd
 E}\frac{m}{E}\frac{dm}{d\varphi}\right)^{\rm l.eq.}\,\delta\varphi\nl
 &=&\int d\Gamma\left(E-\frac{Tm}{E}\frac{dm}{dT}\right)^{\rm l.eq.}
 \widetilde\delta f\,.
\end{eqnarray}
One can easily see that expression (\ref{LLKapusta}) is not zero
if (\ref{delf1}), i.e., the Landau-Lifshitz condition (\ref{LL}),
is not fulfilled. Performing the variable shift (\ref{variable1})
one can find the parameter $\widetilde{b}$ to satisfy condition
(\ref{LL}). Thus, we find
 \be\label{tildeb}
\widetilde{b}=\frac{3c_s^2}{\left\langle\vec{p}^{\,2}\right\rangle}
\,\left\langle\widetilde\tau Q\left(E-\frac{Tm}{E}\frac{\rmd
m}{\rmd T}\right)\right\rangle.
 \ee
 However, we observe that after
the variable shift  the kinetic equation within the initially
assumed  Eq. (\ref{St2}) is no longer fulfilled. Nevertheless, we
can find the new kinetic equation
\begin{eqnarray}
\label{St3}
  {\rm St}\, f = -\widetilde\tau^{-1}\left[\widetilde\delta
  f+\widetilde{b}\frac{E^{\rm l.eq.}}{T}\,f^{\rm l.eq.}(1+f^{\rm l.eq.})
  \,{\vec{\nabla}}\cdot\vec{u}\right]~,
\end{eqnarray}
 being in agreement with the
 performed variable shift and the Landau-Lifshitz condition.

Obviously, the modification (\ref{St3}) compared to the result
(\ref{St2}) does not affect the result (\ref{shear1}) for the
shear viscosity. Combining (\ref{deltaTiiKap}), (\ref{St3}), and
(\ref{tildeb}) one arrives at the result of Ref. \cite{CK10} for
the bulk viscosity:
\begin{eqnarray}\label{genzeta3}
  \zeta &=&\frac{1}{T} \left\langle\widetilde{\tau} Q^2\right\rangle .
  \end{eqnarray}
For $m=\,$const Eq. (\ref{genzeta3}) transforms to (\ref{mconst}),
reproducing the result of Ref. \cite{G85}.

Note that by means of Eq. (\ref{lhs}) we can rewrite (\ref{St3}) as
\begin{eqnarray}\label{St32}
 {\rm St}\, f[\vec{\nabla} \cdot\vec{u}] &=&
 -\frac{\widetilde{\delta}f[\vec{\nabla} \cdot\vec{u}]}{{\tau}_3 (E[\vec{\nabla} \cdot\vec{u}])}
\end{eqnarray}
with
\begin{eqnarray}
 {\tau}_3 (E)=
 \widetilde{\tau}\left(1+\frac{\widetilde{b}
 E^{\rm l.eq.}}{\widetilde{\tau}Q}\right)\,.
\end{eqnarray}
Thus,  the collision term (\ref{St3}) reaches zero for $f=f^{\rm
l.eq.}(E[\vec{\nabla} \cdot\vec{u}])$  given by (\ref{leqdf1}).

Finally, let us estimate the difference between values $\zeta_{\rm
I}$ [Eq. (\ref{genzeta})] and $\zeta_{\rm III}$ [Eq.
(\ref{genzeta3})]. After simple but lengthy algebra for $\tau
=\widetilde{\tau}=\,$const we obtain \be \frac{\zeta_{\rm
I}-\zeta_{\rm III}}{\zeta_{\rm III}} = T\frac{dm}{dT}\frac{
\left\langle\frac{m}{E^2}\right\rangle -
\frac{\left\langle{m}\right\rangle}{
\left\langle{E^2}\right\rangle} \left\langle{1} \right\rangle}
{\left\langle{1}\right\rangle-T\frac{dm}{dT}\left\langle{\frac{m}{E^2}}
\right\rangle}.\nl \ee Then performing numerical evaluations we
arrive at inequalities \be\label{dif} 0\leq \frac{\zeta_{\rm
I}-\zeta_{\rm III}}{\zeta_{\rm III}}\leq
 T\frac{dm}{dT}\frac{
\left\langle\frac{m}{E^2}\right\rangle -
\frac{\left\langle{m}\right\rangle}{
\left\langle{E^2}\right\rangle} \left\langle{1} \right\rangle}
{\left\langle{1}\right\rangle
-\left\langle\frac{m}{E^2}\right\rangle}<0.23 \ee for $
0<T\frac{dm}{dT}<m$, as in the given gluon model, and
 \be
0.15 \frac{Tdm}{mdT}\leq T\frac{dm}{dT}\frac{
\left\langle\frac{m}{E^2}\right\rangle -
\frac{\left\langle{m}\right\rangle}{ \left\langle{E^2}\right\rangle}
\left\langle{1} \right\rangle} {\left\langle{1}\right\rangle}
\leq\frac{\zeta_{\rm I}-\zeta_{\rm III}}{\zeta_{\rm III}}
 \ee
for $\frac{dm}{dT}<0$, as in case of the baryon-less matter
described in the relativistic mean-field hadronic models (see
\cite{KTV09,KTV10}). Thus we see that in our gluon model values of
$\zeta$ evaluated following expressions (\ref{genzeta}) and
(\ref{genzeta3}) deviate from each other by less  than 23\%.

\section{Results of calculations of shear and bulk viscosities}

Analytical expressions for the shear viscosity coincide in all
three ansatze provided one uses the same value of the averaged
relaxation time $\bar{\tau}$.

For the bulk viscosity, the
  three expressions  (\ref{genzeta}),  (\ref{genzeta1}), and
(\ref{genzeta3}) coincide in the limit $m=\,$const but they differ
in a general case. Also, results for $\zeta$  obtained within
ansatz I [Eqs. (\ref{genzeta}) and (\ref{genzetashift})], ansatz
II [Eqs.  (\ref{genzeta1}) and (\ref{genzeta1shift})], and ansatz
III [Eq. (\ref{genzeta3})] approximately coincide in the
nonrelativistic limit. In order to estimate differences in the
bulk viscosities calculated following Eqs. (\ref{genzetashift}),
(\ref{genzeta1shift}), and (\ref{genzeta3}), we further perform
numerical evaluations.

First,  we should choose the value of the averaged relaxation time
$\bar{\tau}$. In Ref. \cite{KTVglue1}  two parametrizations were
used. The first parametrization,
 \be
 \label{Cass}{\bar\tau}^{-1}=a_0 \, {g^2\,T}
\ln ({2c}/{g^2})~
\ee
 with $a_0 =N_c /4\pi \simeq 0.2387$ and a tuning parameter $c$
(where in  \cite{KTVglue1} we have chosen $2c= g^2(T_c)$), is
based on nonperturbative evaluations~\cite{PC05}. The deficiency
of this parametrization is that it does not reproduce the
appropriate perturbative limit for $g^2\ll 1$. The second
parametrization, previously applied in Ref. ~\cite{BKR09},
${\bar{\tau}}^{-1} =a_\eta /(32\pi^2)\, g^4\,T \ln (a_\eta \pi /
g^2)~$ with  a tuning parameter $a_\eta $, allows one to reproduce
an appropriate perturbative limit for $g^2\ll 1$. In recent papers
\cite{BKR10}, the authors slightly modified the latter
parametrization as
 \be \label{Kam}
{\bar{\tau}}^{-1} =a_1 g^4 T\ln(a_2  / g^2)\,,
 \ee
introducing  a  tuning  parameter $a_2$; here
$a_1=0.02587\simeq a_\eta/(32\pi^2)$.
We use $a_2 = g^2(T_c)$ since for $a_2=(\mu^*/T)^2
<g^2(T_c)$ used in Refs. \cite{BKR09,BKR10} the result (\ref{Kam})
becomes negative in some temperature interval above $T_c$.
\begin{figure}[thb]
\includegraphics[width=8.truecm,clip]{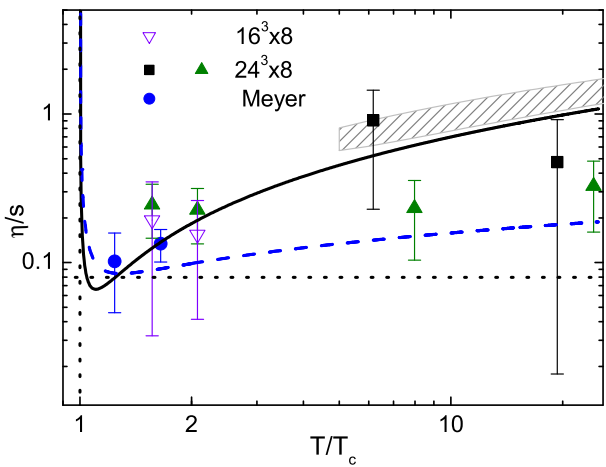}
\caption{(Color online) The  ratio of the gluon shear viscosity to
the entropy density as a function of reduced temperature. Solid
and dashed lines are computed with the averaged relaxation time
obeying Eq. (\ref{Kam}) and
 Eq. (\ref{Cass}), respectively,  with $2c=a_2=g^2(T_c)$.
The horizontal dotted line is the  AdS/CFT bound $\eta/s=1/4\pi$
~\cite{KSS03}. The lattice gauge ${\rm SU}(3)$ data with
$16^3\times8$ and $24^3\times8$ lattice are from Ref.~\cite{SN07}
(triangles and squares) and \cite{Ma07} (filled circles). The
shaded band corresponds to the perturbative result of Ref.
\cite{AMY}. The vertical dotted line shows the transition
temperature.
 }
 \label{etasnew}
\end{figure}

\begin{figure}[htb]
\includegraphics[width=8.truecm,clip]{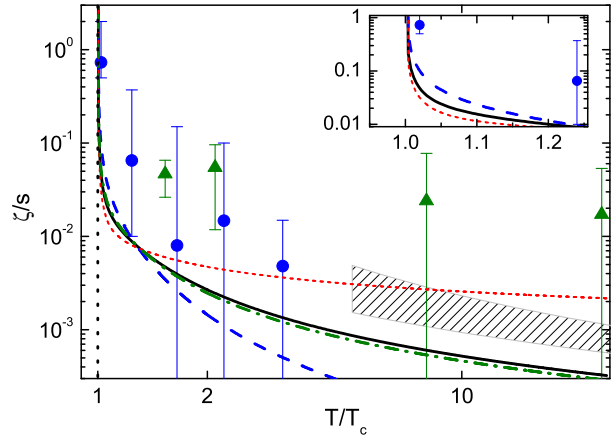}
\caption{(Color online) The  ratio of the gluon bulk viscosity to
the entropy density as a function of temperature. The solid and
long-dashed curves are calculated  following Eq. (\ref{genzeta})
(ansatz I) with relaxation time from (\ref{Kam}) and (\ref{Cass}),
respectively. The short-dashed  curve is the ansatz II result [Eq.
(\ref{genzeta1})] for the relaxation time from Eq. (\ref{Kam}).
The dash-dotted line corresponds to the ansatz III calculations
[Eq. (\ref{genzeta3})] with parametrization (\ref{Kam}).
 Lattice points are taken from \cite{SN07} (triangles)
and~\cite{Me08} (filled circles). The shaded band corresponds to
the perturbative result from Ref. \cite{Arnold1}. The vertical
dotted line shows the transition temperature.}
 \label{zetasnew}
\end{figure}

 The temperature dependence of  the $\eta/s$ and $\zeta/s$
ratios is presented in Figs. \ref{etasnew} and \ref{zetasnew},
respectively. The lattice data \cite{SN07} were obtained using the
lattice entropy from the old paper \cite{BEG95}. We corrected
these data  for $T/T_c>2$ in accordance with the new lattice QCD
result \cite{Pa09}, which resulted in an increase of the lattice
points in Figs. \ref{etasnew} and~\ref{zetasnew} by about 20\%
compared to the original paper \cite{SN07}. As we have noted, all
three ansatze result in the same values of the shear viscosity.
The dashed and solid curves in Fig. \ref{etasnew}, corresponding
to the two parametrizations (\ref{Cass}) and (\ref{Kam}) of the
averaged relaxation time $\bar{\tau}$, have similar trends but
different absolute values. With selected parameters $c$ and $a_2$
in (\ref{Cass}) and (\ref{Kam}) the curves exhibit discontinuity
at the critical temperature $T_c$.
 Using (\ref{Kam}) we can  accurately describe the perturbative
tail at very high temperatures.

In Fig. \ref{zetasnew}, calculations using different expressions for the
bulk viscosity are compared with the lattice result. One should note
that although the presented lattice data are not very conclusive due to
large error bars, nevertheless, one may make some conclusions.
The $\zeta/s$ ratios calculated with the relaxation time
(\ref{Kam}) within ansatze I, II, and III are presented by the
solid, short-dashed, and dash-dotted curves, respectively. The
results for ansatz I [Eq. (\ref{genzeta}), solid curve] and ansatz
III [Eq.(\ref{genzeta3}), dash-dotted curve] are very close to
each other, in agreement with (\ref{dif}), whereas the ansatz II
result [(\ref{genzeta1}), short-dashed curve]  differs from them
significantly and overestimates lattice data in the perturbative
regime. The result (\ref{genzeta1}) of ansatz II with the
relaxation time (\ref{Cass}) (see Fig. 4 of \cite{KTVglue1})
proves to be close to those presented by the solid and dash-dotted
curves. On the other hand, the result (\ref{genzeta}) (long-dashed
curve, ansatz I) with the relaxation time (\ref{Cass})  perhaps
 underestimates the lattice data, whereas in the
nonperturbative regime (at $T\sim T_c$) the long-dashed curve
using the relaxation time (\ref{Cass}) is closer to the data
points than the solid curve using the relaxation time (\ref{Kam}).

\section{Conclusions}
We obtained expressions for the gluon shear and bulk viscosities
within the relaxation time approximation by making use of an
averaged value for the relaxation time $\tau
=\widetilde{\tau}=\bar{\tau}$. Under this assumption, the
expressions for the shear viscosity
 within ansatz I [Eq. (\ref{shear})],  ansatz II,
and ansatz III [Eq. (\ref{shear1}] coincide, which indicates
robustness of the shear viscosity to the ansatz reductions
performed.

In contrast, the expressions for the bulk viscosity,
(\ref{genzeta}), (\ref{genzeta1}), and (\ref{genzeta3}), obtained
within ansatze I, II, and III, respectively, significantly differ.
Our  numerical analysis demonstrates that the results
(\ref{genzeta}) and (\ref{genzeta3}) for gluons those mass
increases with temperature, which use the Landau-Lifshitz
condition (26), are close to each other in the whole range of
temperatures whereas the result (\ref{genzeta1}) using a modified
Landau-Lifshitz condition (27) deviates from them significantly.
One should stress that the result (\ref{genzeta1}) is also
recovered if in making variations one does not vary quantities
which depend on the distribution function only implicitly, such as
$E$ within ansatz I. The later approximation is fully justified
only for nonrelativistic systems.

Among  the results for the bulk viscosity, Eq. (\ref{genzeta})
seems to us most theoretically established. Nevertheless, all
evaluations of the shear and bulk viscosities obtained in the
framework of the relaxation time approximation can be considered
only as rough estimations. In order to perform more established
calculations, one should go beyond the scope of the relaxation
time approximation. However, such calculations are much more
involved than estimations presented in the given work and have not
yet been carried out for systems with strong interactions.

Viscosity coefficients in a weakly coupled scalar field theory at
arbitrary temperature can be  evaluated directly from first
principles without reference to the relaxation time approximation.
This has been done by considering the expansion of the Kubo
formulas in terms of ladder diagrams in the imaginary time
formalism. In a theory with cubic and quartic interactions, the
infinite class of diagrams which contribute to the leading
weak-coupling behavior are identified and summed. The resulting
expression is reduced to a single linear integral equation, which
is shown to be identical to the corresponding result obtained from
a linearized Boltzmann equation similar to those which arise when
the Boltzmann equation is treated in the relaxation time
approximation, as was  first noted in Refs. \cite{Je94,VB02}.
Unfortunately, a similar analysis for more general cases is
unavailable.

\vspace{3mm} \centerline{\bf Acknowledgements}

We are grateful to Pavel Buividovich and Eugeny Kolomeitsev for
useful discussions. This work was supported by RFBR Grant No.
11-02-01538-a and a grant from the European network I3-HP2 Toric.

\end{document}